\documentclass[aps,pra,twocolumn,showpacs]{revtex4-2}
\usepackage{amssymb}
\usepackage{graphicx}
\usepackage{dcolumn}
\usepackage{bm}
\usepackage{amsmath}
\usepackage{soul,color}
\usepackage{textcomp}
\usepackage{times}

\definecolor{cBlue}{RGB}{45,51,180}

\usepackage[colorlinks,linkcolor=red,citecolor=cBlue]{hyperref}

\begin{document}

\title{Floquet geometric squeezing in fast-rotating condensates}
\author{Li Chen$^{1}$}
\email{lchen@sxu.edu.cn}
\author{Fei Zhu$^{1}$}
\author{Yunbo Zhang$^{2}$}
\author{Han Pu$^{3}$}

\affiliation{
$^1${Institute of Theoretical Physics, State Key Laboratory of Quantum Optics and Quantum Optics Devices, Shanxi University, Taiyuan 030006, China}\\
$^2${Key Laboratory of Optical Field Manipulation of Zhejiang Province and Physics Department of Zhejiang Sci-Tech University, Hangzhou 310018, China}\\
$^3${Department of Physics and Astronomy, and Smalley-Curl Institute, Rice University, Houston, TX 77005, USA}
}

\begin{abstract}
Constructing and manipulating quantum states in fast-rotating Bose-Einstein condensates (BEC) has long stood as a significant challenge as the rotating speed approaching the critical velocity.
Although the recent experiment [Science, 372, 1318 (2021)] has realized the geometrically squeezed state of the guiding-center mode, the remaining degree of freedom, the cyclotron mode, remains unsqueezed due to the large energy gap of Landau levels.
To overcome this limitation, in this paper, we propose a Floquet-based state-preparation protocol by periodically driving an anisotropic potential.
This protocol not only facilitates the single cyclotron-mode squeezing, but also enables a two-mode squeezing. 
Such two-mode squeezing offers a richer set of dynamics compared to single-mode squeezing and can achieve wavepacket width well below the lowest Landau level limit.
Our work provides a highly controllable knob for realizing diverse geometrically squeezed states in ultracold quantum gases within the quantum Hall regime.
\end{abstract}

\maketitle
{\it Introduction} --- Quantum simulation of Landau levels using cold atoms holds significance for exploring topological states and discovering novel quantum phases of matter that have no counterpart in electronic materials \cite{Goldman2016, Ozawa2019, Cooper2019,Viefers2008,Cooper2008}. Rotating Bose-Einstein condensates (BECs) provide a viable pathway for such simulations as it can mimic the motion of electrons in a gauge field \cite{Fetter2009, Fetter2007, Madison2000, Shaeer2001, Schweikhard2004, Bretin2004, Fletcher2021, Mukherjee2022, Ho2001, Recati2001, Regnault2003, Sonin2005, Sinha2005, Furukawa2012, Regnault2013}. The corresponding dynamics  encompasses two degrees of freedom – the \textit{cyclotron} mode and the \textit{guiding-center} mode. Particularly, when the BEC is in the quantum Hall regime, i.e., the rotating frequency $\Omega$ approaching the external trapping frequency $\omega$, the effective energy of the guiding-center mode vanishes, leading to extensive level degeneracy. This degeneracy, combined with the non-degenerate cyclotron mode, gives rise to the characteristic Landau levels typically observed in charged particles in two dimensions (2D) subjected to a strong magnetic field \cite{Yoshioka2002,Tong2016}.

Despite various advantages of rotating BECs, precise manipulation of quantum states within the quantum Hall regime is hindered by instabilities \cite{Fetter2009, Sinha2005}. In this case, the centrifugal force exactly counterbalances the confining harmonic potential, rendering the atoms in a flat-land scenario (i.e., the Landau levels) lacking of effective confinement.
Recently, an experiment on geometric squeezing provides an effective approach for quantum control of BECs within the quantum Hall regime \cite{Fletcher2021}. The experiment employed a quasi-2D harmonic potential with weak anisotropy, which effectively provided a transverse Hall drift \cite{Sharma2022, Crépel2024, Shaffer2023}. Under its influence, the quantum fluctuation in the guiding-center phase space was suppressed, akin to degenerate parametric oscillation in quantum optics \cite{Wu1986, Wu1987, Scully1997, Walls2008}, leading to a single-mode (i.e., the guiding-center mode) squeezed state. The real-space density distribution of the BEC becomes an anisotropy Gaussian with a minimal width $\sigma_\text{LLL}$ \cite{Fletcher2021}, the characteristc length of the lowest Landau level. The width $\sigma_\text{LLL}$ arises from the unsqueezed cyclotron mode, with the associated wave function remaining in the ground state of a harmonic oscillator.

In fact, similar to the guiding-center mode, the Hamiltonian realized in the experiment also provides the necessary terms for squeezing the cyclotron mode \cite{Fletcher2021}. However, due to the dominant energy gap of the Landau levels, effective geometric squeezing of the cyclotron mode is unattainable. In other words, to achieve significant squeezing in the cyclotron mode, we need to find a way to overcome this energy gap. Motivated by this question, in this paper, we propose a Floquet-based state-preparation protocol, in which the anisotropy of the trap is periodically modulated. We find that, when the modulation frequency $\nu$ coincides with twice the energy gap of Landau levels, the Floquet effective Hamiltonian can circumvent the aforementioned limitation and efficiently generate squeezing of the cyclotron mode. More importantly, a protocol comprising both direct (DC) and alternating (AC) components allows for the simultaneous squeezing of both the cyclotron and the guiding-center modes, resulting in a two-mode geometrically squeezed state. In real space, the wavepacket width of the two-mode squeezed state decays exponentially and can surpass the limitation of $\sigma_\text{LLL}$. 

{\it Hamiltonian} --- We consider the experimental setup \cite{Fletcher2021}: a quasi-2D BEC being loaded into a magnetic harmonic trap. The trap is rotating along the $z$-direction in angular frequency $\Omega$. In the rotating reference, the system is described by the single-particle Hamiltonian (setting $\hbar = 1$)
\begin{equation}
h_0 = \frac{\mathbf{p}^2}{2m} + V_\text{ext}(\mathbf{r}) - \Omega L_z,
\label{h0}
\end{equation}
where $\mathbf{r}=(x,y)$ and $\mathbf{p}=(p_x,p_y)$, $m$ is the atomic mass, and $L_z = x p_y - y p_x$ denotes the axial angular momentum operator. The external potential $V_\text{ext}$ is given by
\begin{equation}
V_\text{ext}(\mathbf{r}) = \frac{m (1+\varepsilon)\omega^2}{2} x^2 + \frac{m (1-\varepsilon)\omega^2}{2} y^2,
\label{Vext}
\end{equation}
with $\varepsilon \ll 1$ being a small dimensionless parameter characterizing the anisotropy of the trap. Notably, when $\varepsilon \neq 0$, the axial rotational symmetry is broken.

To separate the cyclotron and the guiding-center modes, we perform a unitary transformation $G = \exp(-i\kappa m \omega x y)$ with $\kappa \equiv \varepsilon \omega/2\Omega$. The transformed Hamiltonian reads
\begin{equation}
\begin{aligned}
\tilde{h}_0=&G h_0 G^\dagger = s \left[\frac{\tilde{p}^2}{2 m}+\frac{m \omega^2}{2} \left(\tilde x^2+\tilde y^2\right)\right] \\
&+\kappa \omega\left(\tilde x \tilde p_y+\tilde y \tilde p_x\right)-\Omega \left(\tilde x \tilde p_y-\tilde y \tilde p_x\right),
\end{aligned}
\label{h01}
\end{equation}
where $s \equiv \sqrt{1+\kappa^2}$ and
\begin{equation}
\tilde{\mathbf{r}} = (\tilde x, \tilde y) = s^{1/2} \mathbf{r}\,, \ \ \ \ 
\tilde{\mathbf{p}} = (\tilde{p}_x, \tilde{p}_y) = s^{-1/2} \mathbf{p}\,.
\end{equation}
Then, one can define two sets of independent bosonic modes: the \textit{cyclotron mode} characterized by the ladder operator $\tilde a$ and quadratures $(\tilde \xi, \tilde \eta)$, and the \textit{guiding-center} mode characterized by the operator $\tilde b$ and quadratures $(\tilde X, \tilde Y)$. The specific definitions of the mode operators are given by 
\begin{equation}
\begin{aligned}
\tilde a &= \frac{\tilde \xi + i \tilde \eta}{\sqrt {2} l_B}, \ \ \ \ \ \ \ 
\tilde \xi = \frac{\tilde x}{2} - \frac{\tilde p_y}{2 m \omega}, \ \ \ \ 
\tilde \eta = \frac{\tilde y}{2} + \frac{\tilde p_x}{2 m \omega}, \\
\tilde b &= \frac{\tilde X - i \tilde Y}{\sqrt {2}l_B}, \ \ \ 
\tilde X = \frac{\tilde x}{2} + \frac{\tilde p_y}{2 m \omega}, \ \ \ 
\tilde Y = \frac{\tilde y}{2} - \frac{\tilde p_x}{2 m \omega},
\end{aligned}
\label{opa}
\end{equation}
with $l_B \equiv 1/\sqrt{2m\omega}$ being the magnetic length of the Landau levels (see below). 
Notably, operators belonging to each of the two modes satisfy bosonic commutation relations, i.e.,
\begin{equation}
[\tilde a,\tilde a^\dagger] = [\tilde b,\tilde b^\dagger]=1,\;
[\tilde \xi,\tilde \eta] = -[\tilde X,\tilde Y] = i l_B^2;
\label{commutator1}
\end{equation}
whereas, operators between the two modes mutually commute. In terms of these operators, the Hamiltonian $\tilde{h}_0$ takes a simple form of
\begin{equation}
\tilde h_0 =\tilde \omega_+\tilde{a}^{\dagger} \tilde{a}+\tilde\omega_-\tilde{b}^{\dagger} \tilde{b} - \frac{ \zeta}{2}\left(\tilde{a}^2 -\tilde{b}^2 + \text{h.c.}\right),
\label{h02}
\end{equation}
where $\tilde\omega_\pm \equiv s\omega \pm \Omega$ and $\zeta \equiv \kappa \omega = \varepsilon \omega^2 /2\Omega$. The separation of the cyclotron and the guiding-center modes becomes manifest. Now, $|n_{\tilde a},n_{\tilde b}\rangle$ provides a complete set of basis, where the non-negative integers $n_{\tilde a}$ and $n_{\tilde b}$ are the quantum numbers associated with $\tilde{a}^{\dagger} \tilde{a}$ and $\tilde{b}^{\dagger} \tilde{b}$, respectively.

One immediately notices that the terms $\tilde{a}^2$ and $\tilde{b}^2$ in Hamiltonian~(\ref{h02}), which resemble the parametric conversions in quantum optics \cite{Wu1986,Wu1987}, serve as the basis for the geometric squeezing. The squeezing parameter $\zeta$ is proportional to the trap anisotropy parameter $\varepsilon$. At the critical rotation velocity $\Omega=\omega$, the three key parameters characterizing $\tilde h_0$: $\tilde \omega_+ \approx 2\omega$, $\zeta=\varepsilon \omega/2$ and $\tilde \omega_- \approx \varepsilon^2 \omega/8$ are of zeroth, first and second order in $\varepsilon$, respectively. For a small $\varepsilon$, $\tilde \omega_-$ becomes negligible such that the first two terms of $\tilde h_0$ yield the Landau levels: for a given $n_{\tilde{a}}$, different $n_{\tilde{b}}$ provide massive degeneracy; in contrast, states in adjacent $n_{\tilde{a}}$ exhibit an energy gap $\tilde \omega_+$. Particularly, states $|n_{\tilde a} = 0,n_{\tilde b}\rangle$ are called the lowest Landau levels (LLL). 

\begin{figure}[t]
	 \includegraphics[width=0.46\textwidth]{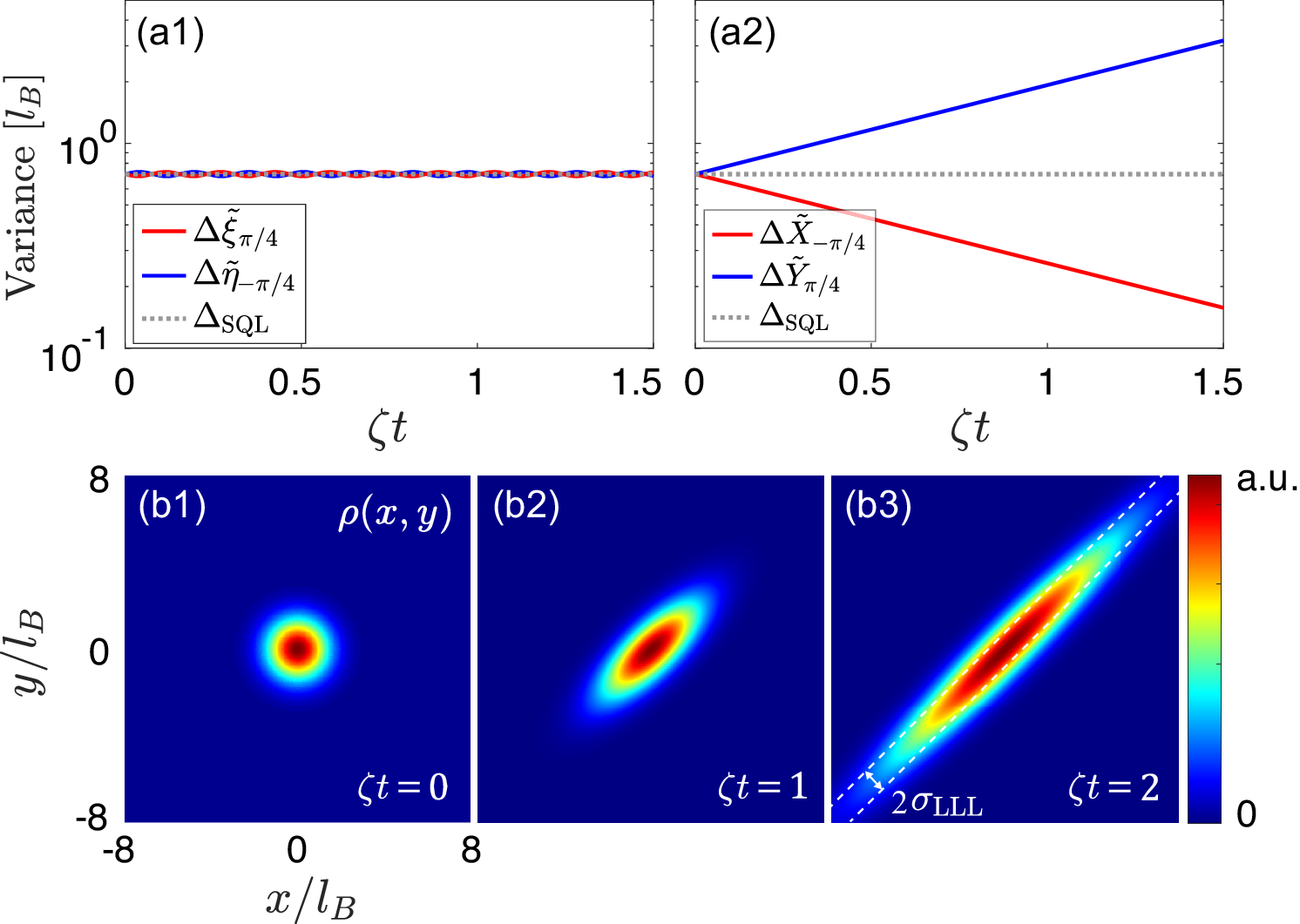}
	\caption{Guiding-center mode squeezing. (a1) and (a2) show the quadrature fluctuations in the $\tilde \xi$-$\tilde \eta$ and $\tilde X$-$\tilde Y$ phase spaces, respectively. (b1)-(b3) display the real-space density distribution $\rho(\mathbf{r},t)$ [in arbitrary units (a.u.)] at selected moments, with white dashed lines indicating $\pm \sigma_\text{LLL} = \pm l_B/\sqrt{2}$. In our calculation, we take $\varepsilon = 0.2$.
	}
	\label{Fig1}
\end{figure}

\begin{figure*}[t]
	 \includegraphics[width=1.00\textwidth]{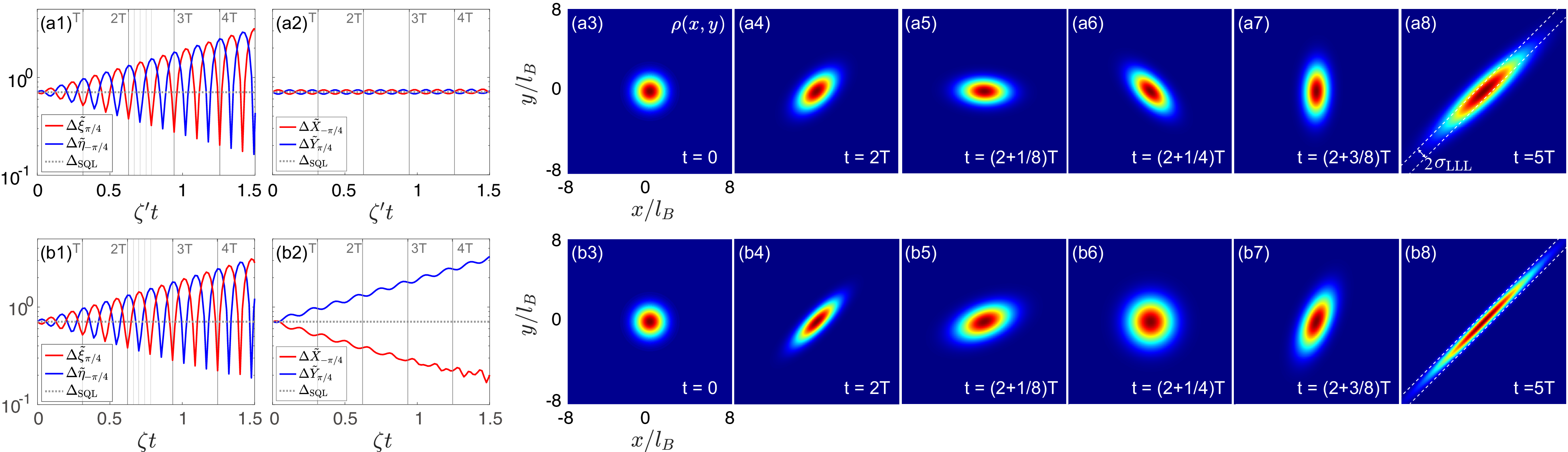}
	\caption{Floqeut geometric squeezing. The upper row (a) illustrates the cyclotron-mode squeezing with $\varepsilon = 0$ and $\varepsilon' = 0.2$; the lower row (b) shows the two-mode squeezing with $\varepsilon = \varepsilon' = 0.2$. In each row, the first two panels show the quadrature fluctuations in the $\tilde \xi$-$\tilde \eta$ and $\tilde X$-$\tilde Y$ phase spaces; vertical lines mark the stroboscopic moments $t = nT$, with $T = 2\tilde{\omega}_+\approx 4\omega$; the remaining six panels display the density distribution $\rho(\mathbf{r},t)$ at selected moments. In the Supplemental Materials (SM) \cite{SM}, we provide animations of $\rho(\mathbf{r}, t)$ for various types of geometric squeezing.
	}
	\label{Fig2}
\end{figure*}

\textit{Guiding-Center Mode Squeezing ---} Consider the following protocol: 1) Prepare the BEC in the ground state of the system with the isotropic irrotational trap (i.e., $\varepsilon=0$ and $\Omega = 0$); 2) Ramp up the rotation frequency $\Omega$, in which the BEC remains in the isotorpic steady state \cite{Note1}; 3) At $t=0$, when the critical condition $\Omega=\omega$ is reached, the trap anisotropy $\varepsilon$ is suddenly turned on and the system starts to evolve under Hamiltonian~(\ref{h02}). Note that the initial non-rotating state is the same as $|n_{\tilde a}=0,n_{\tilde b}=0\rangle$ up to a correction of $O(\varepsilon^2)$. In the ensuing time evolution, the cyclotron mode is dominated by the first term $\tilde \omega_+\tilde{a}^{\dagger} \tilde{a}$, which thus, to a good approximation, remains in the initial state $|n_{\tilde a}=0 \rangle$. In contrast, the time-evolution operator for the guiding-center mode takes the form of the squeezing operator $U_{\tilde b}(t) = \exp(-i \zeta t  \tilde b ^2/2 - \text{h.c.})$ with squeezing angle $-\pi/4$, which transforms the BEC into a single-mode squeezed state, i.e., $U_{\tilde b}(t) |0,0\rangle = |0,S(t) \rangle$. During the squeezing, quantum fluctuations behave as
\begin{equation}
\Delta \tilde X_{-\pi/4} = \Delta_\text{SQL} e^{- \zeta t}, \ \ \ \Delta \tilde Y_{\pi/4} = \Delta_\text{SQL} e^{ \zeta t},
\label{DeltaXt}
\end{equation}
where $\Delta\tilde{X}_{-\pi/4}$ and $\Delta\tilde{Y}_{\pi/4}$ denote the quadrature fluctuations in the $\tilde X$-$\tilde Y$ phase space respectively along the squeezing and anti-squeezing directions, and $\Delta_\text{SQL} = l_B/\sqrt{2}$ is the standard quantum limit (SQL). In the coordinate $x$-$y$ space, the BEC's density distribution can be obtained as \cite{Fletcher2021,SM}
\begin{align}
\rho(\mathbf{r},t) &\approx |\langle {\mathbf{r}}|0,S(t)\rangle|^2 \nonumber \\
& =\frac{e^ {-[1-\tanh (\zeta t)] \frac{( x+ y)^2}{4 l_B^2}-[1+\tanh (\zeta t)] \frac{( x- y)^2}{4 l_B^2}}}{2\pi l_B^2 \cosh (\zeta t)}, 
\label{rhoGuid}
\end{align}
which is a 2D Gaussian independently along directions $(x \pm y)/\sqrt{2}$ and of widths $\sqrt{(1 + e^{\pm 2 \zeta t})/2} l_B$, respectively. In the asymptotic limit $t \rightarrow \infty$, the width along the $-\pi/4$ direction converges to $\sigma_\text{LLL} = l_B/\sqrt{2}$, while that along the $\pi/4$ direction diverges. 

We confirm these results by numerically solving the time-dependent Schr\"{o}dinger equation based on Eqs.~(\ref{h0}) and (\ref{Vext}), with the outcome displayed in Fig.~\ref{Fig1}. Specifically, quadrature fluctuations in the $\tilde \xi$-$\tilde \eta$ and $\tilde X$-$\tilde Y$ phase spaces are presented in panels (a1) and (a2), respectively; the density profiles at selected moments are displayed in panels (b1)-(b3).

{\it Floquet Protocol} --- Now we introduce our Floquet protocol to squeeze the cyclotron mode. The full protocol is similar to what is described above except that the trap anisotropy is periodically modulated as
\begin{equation}
\varepsilon(t) = \varepsilon - 2 \varepsilon' \cos(\nu t),
\label{epsilont}
\end{equation}
where $\varepsilon$ and $\varepsilon'$ are respectively the amplitude of the DC and AC components, and $\nu$ denotes the modulation frequency. The factor 2 is introduced for convenience. Under the same transformation $G$, the Hamiltonian now reads \cite{SM}
\begin{equation}
\begin{aligned}
&\tilde{h}_0(t) = \tilde \omega_+ \tilde{a}^{\dagger} \tilde{a} + \tilde \omega_- \tilde{b}^{\dagger} \tilde{b} - h_\text{ab}(t) \\
 &- \left\{\left[ \frac{\zeta}{2} + \zeta' \cos(\nu t) \right]  \tilde{a}^2- \left[ \frac{\zeta}{2} - \zeta' \cos(\nu t) \right]  \tilde{b}^2 + \text{h.c.}\right\}.
\end{aligned}
\label{h0t}
\end{equation}
Comparing with Eq.~(\ref{h02}), the AC driving provides an alternating squeezing parameter with $\zeta' = \varepsilon' \omega/2 s$, and $h_\text{ab}(t) = 2\zeta' \cos(\nu t)(\tilde a^\dagger \tilde b + \tilde b^\dagger \tilde a)$ denoting the coupling between the two modes.

Taking another unitary transformation $W(t) = e^{i \tilde{\omega}_+ t \tilde{a}^{\dagger} \tilde{a}}$, the Hamiltonian is expressed as
\begin{equation}
\begin{aligned}
&\tilde{h}_0^W(t) = \tilde \omega_- \tilde{b}^{\dagger} \tilde{b} - [2 \zeta' e^{i \tilde{\omega}_+ t} \cos(\nu t) \tilde a^\dagger \tilde b + \text{h.c.}] \\
&- \left\{\left[ \frac{\zeta}{2} + \zeta' \cos(\nu t) \right]  e^{2i \tilde{\omega}_+ t}\tilde{a}^2- \left[ \frac{\zeta}{2} - \zeta' \cos(\nu t) \right]  \tilde{b}^2 + \text{h.c.}\right\}.
\end{aligned}
\label{h0W}
\end{equation}
Now, the term $\tilde{a}^2$, as well as the coupling $\tilde a^\dagger \tilde b$, depends on both $\tilde{\omega}_+$ and $\nu$. We find that, when the modulation frequency is set to $\nu = 2\tilde{\omega}_+ = 2(s\omega + \Omega)$, the Floquet effective Hamiltonian takes the form of
\begin{equation}
\tilde{h}_0^\text{eff} \equiv \frac{1}{T}\int_0^T \tilde{h}_0^W(t) \ dt  = \tilde \omega_- \tilde{b}^{\dagger} \tilde{b} - \left(\frac{ \zeta'}{2}\tilde{a}^2 - \frac{ \zeta}{2} \tilde{b}^2 + \text{h.c.}\right),
\label{h0W2}
\end{equation}
where $T \equiv 2\pi/\tilde{\omega}_+ = 4\pi/\nu$ is the stroboscopic period. The time integral in Eq.~(\ref{h0W2}) kept all the zero-frequency terms in $\tilde{h}_0^W$ but erased all nonzero-frequency terms. 

Equation~(\ref{h0W2}) is a key result of this paper: compared to Eq.~(\ref{h02}), the term $\tilde{\omega}_+ \tilde{a}^\dagger \tilde{a}$ is absent in the effective Hamiltonian $\tilde{h}_0^\text{eff}$, i.e., the Landau-level gap that previously prevented squeezing in the cyclotron mode is now eliminated by the Floquet driving. As a consequence, the term $\sim \tilde{a}^2 + (\tilde{a}^\dagger)^2$ now can dominate the dynamics and generate squeezing in the cyclotron mode.
Furthermore, at the critical rotation with $\Omega=\omega$ and hence $\tilde{\omega}_- \approx 0$, $\tilde{h}_0^\text{eff}$ still allows squeezing of the guiding-center mode. As a consequence, both modes can be squeezed simultaneously, resulting in two-mode geometric squeezing. Below, we will discuss these two cases in detail.

\textit{Cyclotron-Mode Squeezing ---} By turning off the DC component, i.e., setting $\varepsilon = 0$, the $\tilde{b}^2$ term in $\tilde{h}_0^\text{eff}$ vanishes and the guiding-center mode is not squeezed. In the critical case $\Omega=\omega$, we simply have $s = 1$, $\nu = 4\omega$, and $\zeta' = \varepsilon' \omega / 2$. The stroboscopic dynamics at moments $t = nT$ ($n$ being a non-negative integer) is governed by the Floquet evolution operator $U_{\tilde{a}}^n = \exp(i \zeta' nT  \tilde a ^2/2 - \text{h.c.})$, which is also a squeezing operator and drives the cyclotron mode into a squeezed state, i.e., $U_{\tilde a}^n |0,0\rangle = |S,0 \rangle$. The properties of $|S,0 \rangle$ are quite similar to the guiding-center squeezed state $|0,S \rangle$ discussed previously, except that the squeezing now exists in the $\tilde{\xi}$-$\tilde{\eta}$ phase space. The corresponding quadrature fluctuations behave as
\begin{equation}
\Delta \tilde \xi_{\pi/4} = \Delta_\text{SQL} e^{- \zeta' nT}, \ \ \ \Delta \tilde \eta_{-\pi/4} = \Delta_\text{SQL} e^{ \zeta' nT};
\label{DeltaXit}
\end{equation}
The real-space density distribution can be worked as \cite{SM}
\begin{equation}
\rho(\mathbf{r},t = nT)=\frac{e^ {-[1-\tanh (\zeta' t)] \frac{( x+ y)^2}{4 l_B^2}-[1+\tanh (\zeta' t)] \frac{( x- y)^2}{4 l_B^2}}}{2\pi l_B^2 \cosh (\zeta' t)}.
\label{rhoCyc}
\end{equation}
which is also a 2D Gaussian with minimal width along the $-\pi/4$ direction and converging to $\sigma_\text{LLL} = l_B/\sqrt{2}$ as $n\rightarrow \infty$.

The complete dynamics can be obtained by numerically solving the time-dependent Schr\"{o}dinger equation, with results presented in Fig.~\ref{Fig2}(a), where panels (a1) and (a2) show the fluctuations in the $\tilde{\xi}$-$\tilde{\eta}$ and $\tilde{X}$-$\tilde{Y}$ phase spaces, and panels (a3)-(a8) illustrate $\rho(\mathbf{r},t)$ at selected moments. It is shown that the quadrature variance of the $\tilde a$ mode has been considerably squeezed, whereas that in the $\tilde b$ mode remains unsqueezed, as anticipated. At stroboscopic moments $t = n T$ (indicated by thick vertical lines), $\Delta \tilde{\xi}_{\pi/4}$ and $\Delta \tilde{\eta}_{-\pi/4}$ respectively exhibit exponential squeezing and anti-squeezing, confirming the analytical results in Eq.~(\ref{DeltaXit}).

Within a stroboscopic period $T$, quantum fluctuations oscillate, accompanied by the clockwise rotation of the density profile, which can be understood as follows. For any initial time $t_0$, the Floquet Hamiltonian $\tilde{h}_0^{t_0}$ characterizes the physics at moments $t = t_0 + nT$, and the effective Hamiltonian $\tilde{h}_0^\text{eff}$ shown in Eq.~(\ref{h0W2}) represents the specific case for $t_0=0$. It is straightforward to show that \cite{SM}
\begin{equation}
\tilde{h}_0^{t_0} = \tilde \omega_- \tilde{b}^{\dagger} \tilde{b} - \left[\frac{ \zeta'}{2} (e^{-i \varphi/2}\tilde{a})^2 - \frac{ \zeta}{2} \tilde{b}^2 + \text{h.c.}\right],
\label{h0t0}
\end{equation}
implying that the squeezing angle in the $\tilde{\xi}$-$\tilde{\eta}$ phase space is altered by $\varphi/2 = -\nu t_0/2$. Particularly for $t_0$ being odd multiples of $T/4$, $\varphi = \pm\pi$ [mod $2\pi$] (equivalently $\zeta' \rightarrow -\zeta'$ for $\tilde{h}_0^\text{eff}$), which leads to a swap between squeezing and anti-squeezing directions comparing to the case of $t_0=\varphi=0$. This also explains the alternating of the long and short axis of the density distribution $\rho(\mathbf{r})$ [see Eq.~(\ref{rhoCyc}) and Fig.~\ref{Fig2}(a6)].

\textit{Two-Mode Squeezing ---} We are now ready to discuss the two-mode squeezing protocol where $\varepsilon(t)$ includes both the DC and AC components.
Here, we specifically examine the case of $\zeta' = \zeta$, which can be satisfied by setting $\varepsilon' = \varepsilon$ and $\Omega = (1/2+\sqrt{1+\varepsilon^2}/2)^{1/2}\omega \approx (1 + \varepsilon^2/8)\omega$. A more general discussion for $\zeta' \neq \zeta$ can be found in the SM \cite{SM}. In the current situation, the stroboscopic evolution $U^n = \exp[i \zeta nT (\tilde{a}^2/2 - \tilde{b}^2/2) - \text{h.c.}]$ at moments $t=nT$ is a squeezing operator for both the $\tilde{a}$ and $\tilde{b}$ modes, leading to the two-mode squeezed state $U^n |0,0\rangle = |S,S\rangle$.

The stroboscopic dynamics manifests that quantum fluctuations in both phase spaces scale exponentially, following Eqs.~(\ref{DeltaXt}) and (\ref{DeltaXit}), as numerically confirmed by Figs.~\ref{Fig2}(b1) and (b2). The density distribution of $|S,S\rangle$ is given by \cite{SM}
\begin{equation}
\rho(\mathbf{r},t\!=\!nT)=\frac{1}{2\pi l_B^2}\,{\exp \!\left[- \frac{(x+y)^2}{4 l_B^2 \,e^{2 \zeta t}}\!- \frac{(x-y)^2}{4 l_B^2 \,e^{-2 \zeta t}}\right]},
\label{rho2}
\end{equation}
with minimal width along the $-\pi/4$ direction being $e^{-\zeta n T} l_B$. Notably, in contrast to the single-mode squeezing cases presented in Eqs.~(\ref{rhoGuid}) and (\ref{rhoCyc}) where the minimal width asymptotically saturates to $\sigma_\text{LLL}$, here the minimal width exponentially decreases as $n$ increases and can fall below $\sigma_\text{LLL}$. 
Note that although Eq.~(\ref{rho2}) indicates that the minimal width tends to zero for large $n$, this result is obtained under the effective Floquet Hamiltonian $\tilde{h}_0^\text{eff}$ where high-order corrections are neglected. A calculation based on high-frequency expansion shows that the next-order corrections are $\tilde{a}^\dagger \tilde{a}$ and $\tilde{b}^\dagger \tilde{b}$ with strength $\propto \zeta^2/\tilde{\omega}_+$ \cite{SM}. Consequently, the squeezing behaviors will be limited to to a time scale $\sim \omega/\zeta^2$ which prevents the width going all the way to zero. Nevertheless, the statement that the minimal width can fall below $\sigma_\text{LLL}$ is robust as confirmed by our numerical simulation using the original time-dependent Hamiltonian and illustrated in Fig.~\ref{Fig2}(b8).

Furthermore, the dynamics at quarter-periods $t = nT/4$ ($n$ being odd) manifest isotropic density profiles, as shown in Figs.~\ref{Fig2}(b6). Again, the Floquet Hamiltonian $\tilde{h}_0^{t_0}$ is now equivalent to $\tilde{h}_0^\text{eff}$ subject to $\zeta' \rightarrow -\zeta'$, based on which we can obtain \cite{SM}
\begin{equation}
\rho(\mathbf{r},t=nT/4)=\frac{1}{2 \pi l_B^2 \cosh (2 \zeta t)}\exp\left[-\frac{x^2+y^2}{2 l_B^2\cosh (2 \zeta t)}\right],
\end{equation}
which is a 2D isotropic Gaussian, with width $\sqrt{\cosh (2 \zeta t)} l_B$ monotonically increasing in $t$ and scaling exponentially $e^{\zeta t}l_B$ for $t \gg 1/2\zeta$. 

We additionally remark that, although our discussion above has assumed a small anisotropy ($\varepsilon \ll 1$), numerical calculation \cite{SM} shows that the results remain valid for a sizable anisotropy.

\textit{Interacting BECs ---} So far we have ignored inter-atomic interactions. The case would become more complicated when atomic collisions are included, with the system now being described by the Gross-Pitaevskii equation (GPE)
\begin{equation}
i \dot \psi=\left(h_0+g|\psi|^2\right) \psi,
\end{equation}
where $\psi(\mathbf{r},t)$ is the mean-field wave function, and $g = \sqrt{8\pi \omega_z /m} a_s$ denotes the two-body interaction strength, with $a_s$ the reduced s-wave scattering length in 2D and $\omega_z$ the longitudinal trapping frequency.

\begin{figure}[t]
	 \includegraphics[width=0.46\textwidth]{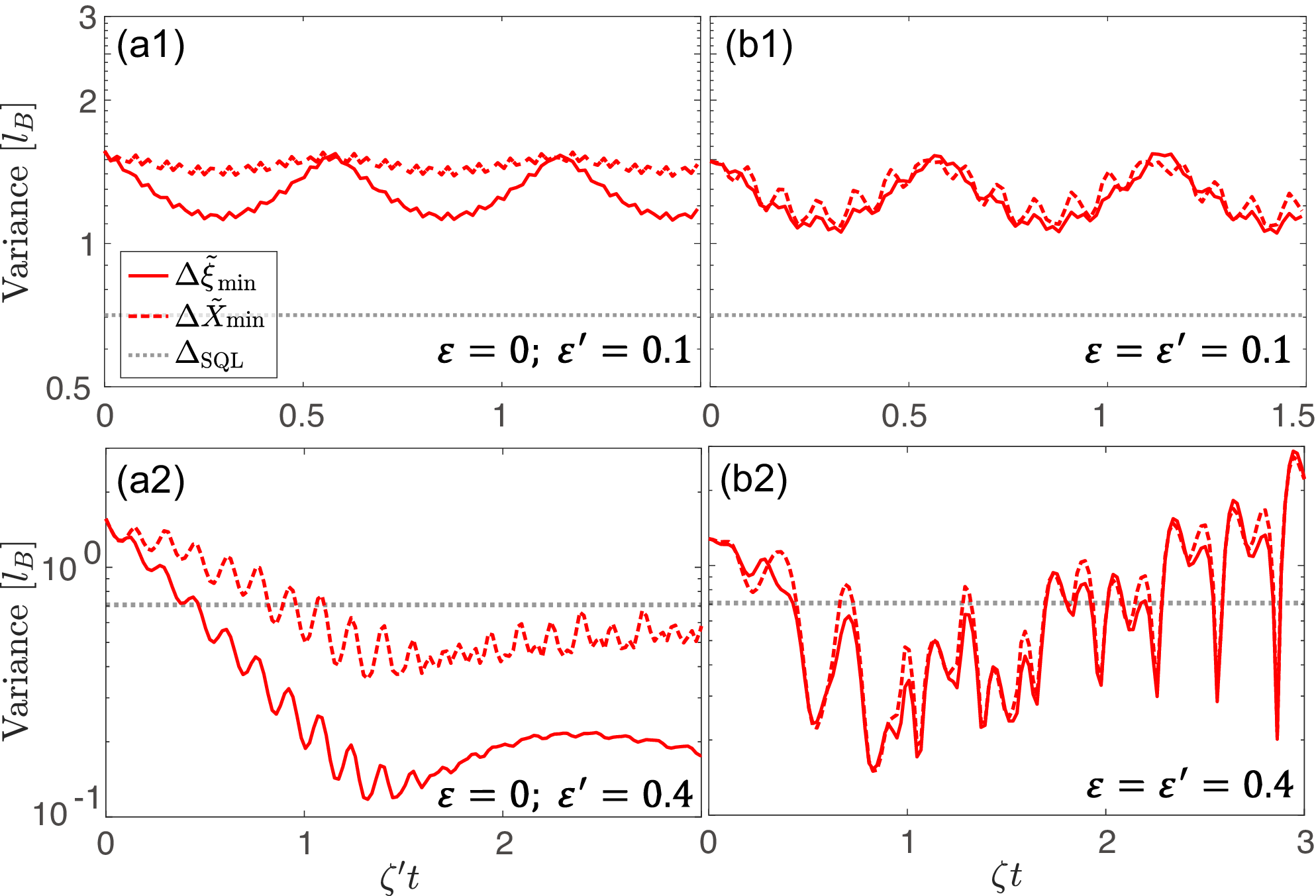}
	\caption{Dynamics of $\Delta \tilde \xi_\text{min}$ (solid lines) and $\Delta \tilde X_\text{min}$ (dashed lines) for interacting BECs. (a1) and (b1) correspond to the cases of the cyclotron-mode and the two-mode squeezings, respectively, with $(\varepsilon, \varepsilon') \sim 0.1$. (a2) and (b2) present the results for $(\varepsilon, \varepsilon') \sim 0.4$.  See the SM \cite{SM} for animations of $\rho(\mathbf{r}, t)$.
	}
	\label{Fig3}
\end{figure}

For sufficiently small $g$, the single-particle physics presented above remains qualitatively unchanged.
However, when the BEC operates within the Thomas-Fermi (TF) regime, i.e., the interaction energy significantly exceeds the kinetic energy, the $g$ term begins to markedly influence the squeezing dynamics.
To illustrate the TF case, we implement our protocol by numerically propagating the GPE, with the results for the cyclotron-mode and the two-mode squeezing being respectively shown in Figs.~\ref{Fig3}(a) and (b). 
The parameters used are close to those in the experiment \cite{Fletcher2021}, i.e., considering $N = 5 \times 10^4$ Na atoms with $a_s \approx 63 a_0$ ($a_0$ being the Bohr radius); the trapping frequencies are $\omega = 88.6 \times (2\pi)$ Hz and $\omega_z = \sqrt{8} \omega$.
In this case, the BEC enters the deep Thomas-Fermi regime, with the wavepacket width being much larger than the harmonic oscillator length (see the animations in Supplemental Material \cite{SM}).
The initial state is the irrotational ground state of the interacting BEC at $\Omega = 0$ and $\varepsilon(t)=0$, which remains stable as $\Omega$ is linearly ramped up to the critical value $\omega$ \cite{Recati2001, Note1}. Then, the system begins to evolve under $\varepsilon(t)\neq 0$.

For the interacting BEC, the squeezing/anti-squeezing direction may not be exactly along $\pm\pi/4$. Hence, we characterize the squeezing by $\Delta\tilde{\xi}_\text{min}$ and $\Delta\tilde{X}_\text{min}$, respectively denoting the minimum quantum fluctuations in $\tilde{\xi}$-$\tilde{\eta}$ and $\tilde{X}$-$\tilde{Y}$ phase spaces. Figs.~\ref{Fig3}(a1) and (b1) present the minimum fluctuations for the cases of small anisotropy $(\varepsilon, \varepsilon') \sim 0.1$. 
The results indicate that neither the single cyclotron mode nor the two-mode can be squeezed effectively, manifested by the periodic oscillations of $\Delta\tilde{\xi}_\text{min}$ and $\Delta\tilde{X}_\text{min}$. The oscillation period $T \approx 0.56 \omega^{-1}$ is insensitive to $g$ when the BEC enters the TF regime. These phenomena imply that the interacting BEC is in a near-equilibrium state, exhibiting certain collective oscillations.

We further find that increasing the anisotropy helps disrupt the periodicity and yields considerable squeezing. Figs.~\ref{Fig2}(a2) and (b2) display the squeezing dynamics for $(\varepsilon, \varepsilon') \sim 0.4$, with all other parameters remaining unchanged. Both scenarios can yield squeezings $\ge -10 \log_{10}(0.15/\Delta_\text{SQL}) \gtrsim 6.7$dB. In real space, $\rho(\mathbf{r}, t)$ exhibits behaviors qualitatively similar to those of the non-interacting cases shown in Fig.~\ref{Fig2}: for $\tilde{a}$-mode squeezing, $\rho(\mathbf{r}, t)$ is elongated during the rotation process, whereas for two-mode squeezing, $\rho(\mathbf{r}, t)$ alternates between isotropic and anisotropic, accompanied by an increase in amplitude.

\textit{Conclusion} --- We have introduced a Floquet protocol by periodically modulating the anisotropy of the trapping potential, resulting in squeezing of both the guiding-center and the cyclotron modes in a rotating BEC. Such two-mode squeezing exhibits a richer set of dynamics in comparison to the one-mode squeezing previously shown and can achieve wavepacket width below the lowest Landau level limit. We also demonstrated the protocol's effectiveness in interacting BECs for relatively large anisotropy. Our work provides a versatile tool for realizing diverse geometrically squeezed states in rotating quantum gases, offering prospect for experimental realization within current experimental capabilities.

\begin{acknowledgments}
We would like to thank Jing Zhang for the valuable discussion. L. C. acknowledges support from the NSF of China (Grants No. 12174236) and from the fund for the Shanxi 1331 Project;
Y. Z. acknowledges support from the NSF of China (Grant No. 12074340) and the Science Foundation of Zhejiang Sci-Tech University (Grant No. 20062098-Y.);
H. P. acknowledges support from the US NSF and the Welch Foundation (Grant No. C-1669).
\end{acknowledgments}

\end{document}

\title[Short title for running header]{Supplemental Materials:
Floquet geometric squeezing in fast-rotating condensates}
\author{Li Chen$^{1}$}
\email{lchen@sxu.edu.cn}
\author{Fei Zhu$^{1}$}
\author{Yunbo Zhang$^{2}$}
\author{Han Pu$^{3}$}

\affiliation{
$^1${Institute of Theoretical Physics, State Key Laboratory of Quantum Optics and Quantum Optics Devices, Shanxi University, Taiyuan 030006, China}\\
$^2${Key Laboratory of Optical Field Manipulation of Zhejiang Province and Physics Department of Zhejiang Sci-Tech University, Hangzhou 310018, China}\\
$^3${Department of Physics and Astronomy, and Smalley-Curl Institute, Rice University, Houston, TX 77005, USA}
}

\begin{abstract}
In this Supplemental Materials (SM), we provide calculation details for the results presented in the main text. The SM is primarily divided into the following sections: Sec.~\ref{SecI} elaborates on the derivation of the Floquet Hamiltonian. Sec.~\ref{SecII} offers detailed calculations related to the real-space wave functions of various geometrically squeezed states. Sec.~\ref{SecIII} discusses the application of the Floquet protocol for systems with large values of trap anisotropy $\varepsilon(t)$. Sec.~\ref{SecIV} presents the high-order corrections to the Floquet Hamiltonian.
\end{abstract}

\maketitle

\affiliation{
		$^{1}${Institute of Theoretical Physics, State Key Laboratory of Quantum Optics and Quantum Optics Devices, Shanxi University, Taiyuan 030006, China} \\
		$^{2}${School of Physics, Sun Yat-Sen University, Guangzhou 510275, China}
	}

\section{Floquet Hamiltonian}
\label{SecI} 

We derive the Floquet effective Hamiltonian based on Eqs.~(1), (2), and (10) of the main text. Under the transformation $G = e^{-i\kappa m \omega x y}$ (with $\kappa = \varepsilon \omega/2\Omega$), we have
\begin{equation}
\begin{gathered}
\tilde{h}_0(t)=G h_0(t) G^{\dagger} \\
\begin{aligned}
&= \frac{p_x^2 + p_y^2}{2 m}+\frac{m \omega^2}{2} \left(1+\kappa^2\right)\left(x^2+y^2\right)-\Omega\left(x p_y-y p_x\right)+\kappa \omega\left(x p_y+y p_x\right) - \varepsilon^{\prime} m \omega^2  \cos (\nu t)\left(x^2-y^2\right) \\
&= \sqrt{1+\kappa^2}\left[\frac{\tilde p_x^2 + \tilde p_y^2}{2 m}+\frac{m \omega^2}{2} \left(\tilde x^2+\tilde y^2\right)\right]-\Omega\left(\tilde x \tilde p_y- \tilde y \tilde p_x\right)+\kappa \omega\left(\tilde x \tilde p_y+\tilde y \tilde p_x\right) - \frac{\varepsilon^{\prime} m \omega^2}{\sqrt{1+\kappa^2}} \cos (\nu t)\left(x^2-y^2\right) \\
&= s \left(\frac{\tilde{\mathbf{p}}^2}{2 m}+\frac{m \omega^2}{2} \tilde{\mathbf{r}}^2\right)-\Omega\left(\tilde x \tilde p_y- \tilde y \tilde p_x\right)+\kappa \omega\left(\tilde x \tilde p_y+\tilde y \tilde p_x\right) - \frac{\varepsilon^{\prime} m \omega^2}{s} \cos (\nu t)\left(\tilde x^2-\tilde y^2\right),
\end{aligned}
\end{gathered}
\tag{S1}  
\label{h0t}
\end{equation}
where $s = \sqrt{1+\kappa^2}$ and the relationship between $({\mathbf{r}},{\mathbf{p}})$ and $(\tilde{\mathbf{r}},\tilde{\mathbf{p}})$ has been shown in Eq.~(4) of the main text. For a small anisotropy $\varepsilon$, $s \approx 1$, $\mathbf{\tilde r}\approx \mathbf{r}$ and $\mathbf{\tilde p}\approx \mathbf{p}$, up to corrections in the order of $O(\varepsilon^2)$. Using the operators of cyclotron and guiding-center modes defined in Eq.~(5) of the main text, Eq.~(\ref{h0t}) can be re-expressed as
\begin{equation}
\begin{aligned}
\tilde{h}_0 (t) & =m \omega \left\{\left(s \omega+\Omega\right)(\tilde{\xi}^2+\tilde{\eta}^2)+\left(s \omega-\Omega\right)(\tilde{X}^2+\tilde{Y}^2) + \kappa \omega \left(\tilde{\eta}^2-\tilde{\xi}^2+\tilde{X}^2-\tilde{Y}^2\right) - \frac{\varepsilon^{\prime} \omega}{s} \cos (\nu t)\left[(\tilde \xi + \tilde X)^2 - (\tilde \eta + \tilde Y)^2\right] \right\} \\ 
& =\left(s \omega+\Omega\right)\left(\tilde{a}^{\dagger} \tilde{a}+\frac{1}{2}\right)+\left(s \omega-\Omega\right)\left(\tilde{b}^{\dagger} \tilde{b}+\frac{1}{2}\right) -\frac{\kappa \omega}{2}\left[(\tilde{a}^{\dagger})^2+\tilde{a}^2-(\tilde{b}^{\dagger})^2-\tilde{b}^2\right] - \frac{\varepsilon' \omega}{2s} \cos (\nu t) \left(\tilde{a}^2 + \tilde{b}^2 + 2\tilde{a}^{\dagger}\tilde{b} + \text{h.c.}\right) \\
& = \tilde \omega _+\left(\tilde{a}^{\dagger} \tilde{a}+\frac{1}{2}\right)+\tilde \omega _-\left(\tilde{b}^{\dagger} \tilde{b}+\frac{1}{2}\right) - \left\{\left[ \frac{\zeta}{2} + \zeta' \cos(\nu t) \right]  \tilde{a}^2- \left[ \frac{\zeta}{2} - \zeta' \cos(\nu t) \right]  \tilde{b}^2 + \text{h.c.}\right\} + 2\zeta' \cos(\nu t)(\tilde a^\dagger \tilde b + \tilde b^\dagger \tilde a),
\end{aligned}
\tag{S2}  
\label{h0t2}
\end{equation}
which reproduces the Eq.~(11) of the main text, with $\tilde\omega_\pm = s\omega \pm \Omega$, $\zeta = \kappa \omega = \varepsilon \omega^2 /2\Omega$ and $\zeta' = \varepsilon' \omega/2 s$.

Then, we introduce the time-dependent transformation $W(t) = e^{i \tilde{\omega}_+ t \tilde{a}^{\dagger} \tilde{a}}$, which commutes with the operators associated with the guiding-center mode, while modifies those related to the cyclotron mode into
\begin{equation}
\begin{aligned}
W(t) \tilde a W^{\dagger}(t)&=e^{i \tilde \omega_+ t} \tilde a, \\
W(t) \tilde a^{\dagger} W^{\dagger}(t)&=e^{-i \tilde \omega_+ t} \tilde a^{\dagger}.
\end{aligned}
\tag{S3}  
\label{Wa}
\end{equation}
Consequently, we obtain the Hamiltonian as
\begin{equation}
\begin{gathered}
\tilde{h}_0^W(t) = W(t) \tilde{h}_0 (t) W^\dagger (t) + i \dot{W}(t) W^\dagger(t) \\
\begin{aligned}
&=  -\left[ \frac{\zeta}{2} + \zeta' \cos(\nu t) \right]\left[e^{-2 i  \tilde\omega_+ t} (\tilde{a}^{\dagger})^2+e^{2 i \tilde \omega_+ t} \tilde{a}^2\right]+ \left[ \frac{\zeta}{2} - \zeta' \cos(\nu t) \right] \left[(\tilde{b}^{\dagger})^2+\tilde{b}^2\right]+2\zeta' \cos (\nu t)\left(e^{-i \tilde \omega_+ t} \tilde{a}^{\dagger} \tilde b+e^{i \tilde \omega_+ t} \tilde{a} \tilde b^{\dagger}\right) \\
&= -\left[\frac{\zeta }{2}e^{2i \tilde \omega_+ t}+\frac{\zeta'}{2} \left(e^{i (\nu+2 \tilde\omega_+)t}+e^{-i (\nu-2 \tilde\omega_+)t}\right)\right] \tilde{a}^2 + \left[ \frac{\zeta}{2} - \frac{\zeta'}{2} \left(e^{i \nu t}+e^{-i \nu t}\right) \right] \tilde{b}^2+2\zeta'\tilde{a}^{\dagger} \tilde b\left(e^{-i (\nu-\tilde \omega_+)t}+e^{i (\nu-\tilde \omega_+)t}\right)+ \text{h.c.}
\end{aligned}
\end{gathered}
\tag{S4}  
\label{h0Wt}
\end{equation}
In the third line, we performed the frequency expansion of all cosine functions. 
One can immediately observe in Eq.~(\ref{h0Wt}) that, the free term $\tilde \omega_+ \tilde{a}^{\dagger} \tilde{a}$ vanishes, accompanied by all the other terms involving $\tilde{a}$ and  $\tilde{a}^\dagger$ being explicitly dependent on both $\tilde{\omega}+$ and the Floquet frequency $\nu$. To be more specific, the square term $\tilde{a}^2$ [the first term in Eq.~(\ref{h0Wt})] incorporates frequencies of $\nu \pm 2 \tilde{\omega}_+$ and $\tilde{\omega}_+$; $\tilde{b}^2$ term includes frequencies $\pm \nu$ and the zero frequency; the coupling terms between the two modes (the third term) involve frequencies $\nu \pm \tilde{\omega}_+$.
When the Floquet frequency is resonant to twice of the Landau energy gap, i.e., $\nu = 2 \tilde{\omega}_+$, Eq.~(\ref{h0Wt}) behaves as
\begin{equation}
\tilde{h}_0^W(t) = -\left[\frac{\zeta }{2}e^{2i \tilde \omega_+ t}+\frac{\zeta'}{2} \left(e^{4 i \tilde\omega_+ t}+1\right)\right] \tilde{a}^2 + \left[ \frac{\zeta}{2} - \frac{\zeta'}{2} \left(e^{2i \tilde\omega_+ t}+e^{-2i \tilde\omega_+ t}\right) \right] \tilde{b}^2+2\zeta'\tilde{a}^{\dagger} \tilde b\left(e^{i \tilde \omega_+t}+e^{-i\tilde \omega_+ t}\right)+ \text{h.c.}.
\tag{S5}
\label{h0Wt2}
\end{equation}
Then, the Floquet effective Hamiltonian can be derived as
\begin{equation}
\tilde{h}_0^\text{eff} = \frac{1}{T}\int_{0}^{T} \tilde{h}_0^W(t ) \ dt = \tilde \omega_- \tilde{b}^{\dagger} \tilde{b} - \frac{\zeta'}{2} \left[(\tilde{a}^{\dagger})^2 + \tilde{a}^2 \right] + \frac{\zeta}{2} \left[(\tilde{b}^{\dagger})^2 + \tilde{b}^2 \right],
\tag{S6}
\end{equation}
which reproduces Eq.~(13) of the main text. Note that the stroboscopic period $T = 2\pi / \tilde \omega_+$ is twice the Floquet period $2\pi/\nu = \pi/ \tilde \omega_+$, because the frequency of the coupling term $\tilde{a}^{\dagger} \tilde b$ manifests as $\nu - \tilde{\omega}_+ = \tilde{\omega}_+$ and $\nu + \tilde{\omega}_+ = 3\tilde{\omega}_+$. Over the period $T$, the net contribution of this term vanishes. 

The Floquet Hamiltonian $\tilde{h}_0^{t_0}$ depicts the stroboscopic dynamics of $\tilde{h}_0(t)$ at moments $t = t_0 + nT$, with $t_0$ the initial time \cite{Eckardt2015}. This is equivalent to the dynamics of $\tilde{h}(t') = \tilde{h}(t - t_0)$ at $t' = nT$, with $t' = t - t_0$. Following procedures similar to those outlined in Eq.~(\ref{Wa})-(\ref{h0Wt2}), one can derive the Hamiltonian in the $W$-frame as
\begin{equation}
\begin{aligned}
\tilde{h}_0^{t_0,W}(t) = &-\left[\frac{\zeta }{2}e^{2i \tilde \omega_+ t}+\frac{\zeta'}{2} \left(e^{i (4 \tilde\omega_+ t - 2 \tilde \omega_+ t_0)}+e^{2i \tilde \omega_+ t_0}\right)\right] \tilde{a}^2 + \left[ \frac{\zeta}{2} - \frac{\zeta'}{2} \left(e^{2i \tilde\omega_+ (t-t_0)}+e^{-2i \tilde\omega_+ (t-t_0)}\right) \right] \tilde{b}^2 \\
&+2\zeta'\tilde{a}^{\dagger} \tilde b\left(e^{i \tilde \omega_+(t-t_0)}+e^{-i\tilde \omega_+ (t-t_0)}\right)+ \text{h.c.},
\end{aligned}
\tag{S7}
\end{equation}
 which is similar to Eq.~(\ref{h0Wt2}) except for the additional $t_0$-dependent phase factor $\pm\nu t_0/2 = \pm\tilde\omega_+ t_0$ for the operator $\tilde{a}$. After the time average, we eventually obtain the Floquet Hamiltonian 
\begin{equation}
\tilde{h}_0^{t_0} = \frac{1}{T}\int_{t_0}^{t_0+T} \tilde{h}_0^{t_0,W}(t) \ dt = \tilde \omega_- \tilde{b}^{\dagger} \tilde{b} - \frac{\zeta'}{2} \left[(e^{i \frac{\varphi}{2}}\tilde{a}^{\dagger})^2 + (e^{-i \frac{\varphi}{2}}\tilde{a})^2 \right] + \frac{\zeta}{2} \left[(\tilde{b}^{\dagger})^2 + \tilde{b}^2 \right],
\tag{S8}
\end{equation}
as presented in Eq.~(16) of the main text, with $\varphi/2 = -\nu t_0/2 = -\tilde\omega_+ t_0$ defined as the altering of the squeezing angle in the $\tilde \xi$-$\tilde \eta$ phase space.

\section{Wave Functions in Coordinate Space}
\label{SecII}

In this section, we will demonstrate the derivation of the real-space wave functions of various geometrically squeezed states, including the single-mode squeezed states of the guiding-center mode and the cyclotron mode, as well as the general case of the two-mode squeezed state. For calculation convenience, we adopt the length unit $1/\sqrt{m\omega}$ and the energy unit $\omega$ of the isotropic harmonic oscillator to make all quantities dimensionless. In this unit, the magnetic length $l_B$ is equal to $1/\sqrt{2}$. 

Our calculations will rely on operators in a spherical basis, defined as \cite{Tong2016}
\begin{equation}
\begin{aligned}
z &= \tilde x + i\tilde y, \ \ \ 
\partial = \frac{1}{2}(i \tilde p_x + \tilde p_y);\\
 \bar{z} &= \tilde x - i\tilde y,\ \ \ 
\bar{\partial} = \frac{1}{2}(-i \tilde p_x + \tilde p_y),
\end{aligned}
\tag{S9}
\end{equation}
which satisfy the following commutation relations:
\begin{equation}
\begin{aligned}
[z,\partial] &= -1, \ \ \  [\bar{z},\bar{\partial}] = 1; \\
[z,\bar{z}] &= [\partial,\bar{\partial}] = [z,\bar{\partial}] = [\bar{z},\partial] = 0.
\end{aligned}
\tag{S10}
\label{commuz}
\end{equation}
In such a basis, the operators for the $\tilde{a}$ and $\tilde{b}$ modes can be re-expressed as
\begin{equation}
\begin{aligned}
\tilde a &= \frac{z}{2} - \bar{\partial} , \ \ \ 
\tilde a^\dagger =  \frac{\bar{z}}{2} - \partial; \\
\tilde b &= \frac{\bar{z}}{2} + \partial, \ \ \ 
\tilde b^\dagger = \frac{z}{2} + \bar{\partial}.
\end{aligned}
\tag{S11}
\label{abz}
\end{equation}
The corresponding inverse representation is given by
\begin{equation}
\begin{aligned}
z &= \tilde b^\dagger + \tilde a, \ \ \ 
\partial = \frac{1}{2}\left( \tilde b - \tilde a^\dagger \right); \\
\bar z &= \tilde b + \tilde a^\dagger , \ \ \ 
\bar \partial = \frac{1}{2}\left( \tilde b^\dagger - \tilde a \right).
\end{aligned}
\tag{S12}
\label{zDz}
\end{equation}
It is also straightforward to evaluate the commutators between the spherical-basis operators and the mode operators as
\begin{equation}
\begin{aligned}
[z,\tilde a] &= [\bar z,\tilde b] = [\partial,\tilde a^\dagger] = [\bar \partial,\tilde b^\dagger] = 0; \\
[z,\tilde a^\dagger] &= [\bar z,\tilde b^\dagger]= -[\bar z,\tilde a]= -[z,\tilde b] = 1; \\
[\partial,\tilde a] &= -[\bar \partial,\tilde b] = -[\bar\partial,\tilde a^\dagger] = [ \partial,\tilde b^\dagger] = \frac{1}{2}.
\end{aligned}
\tag{S13}
\label{zDzCommut}
\end{equation}

\begin{itemize}
\item \textbf{Guiding-center mode squeezing.} The real-space wave function of the guiding-center $\tilde b$-mode squeezed state can be derived as
\begin{equation}
\begin{aligned}
\langle \tilde x,\tilde y| 0,S(t)\rangle &= \langle \tilde x,\tilde y| e^{-i \frac{\zeta t}{2} [(\tilde b^\dagger)^2  + \tilde b^2]} |0,0 \rangle = \langle \tilde x,\tilde y| \frac{1}{\sqrt{\cosh(\zeta t)}} e^{-\frac{i}{2}\tanh(\zeta t) (\tilde b^\dagger)^2 } |0,0 \rangle \\
&=  \langle \tilde x,\tilde y|\frac{1}{\sqrt{\cosh(\zeta t)}} e^{\gamma_b(t) {\left(z- \tilde a\right)^2} } |0,0 \rangle  = \langle \tilde x,\tilde y|\frac{1}{\sqrt{\cosh(\zeta t)}} e^{\gamma_b(t) z^2 } |0,0 \rangle \\
&= \frac{1}{\sqrt{\cosh(\zeta t)}} e^{\gamma_b(t) z^2 } \langle \tilde x,\tilde y|0,0\rangle \\
& = \frac{1}{\sqrt{\pi\cosh(\zeta t)}} e^{\gamma_b(t) z^2 } e^{-\frac{|z|^2}{2}} \\
&= \frac{1}{\sqrt{\pi\cosh(\zeta t)}} e^{\gamma_b(t) (\tilde x+i\tilde y)^2 } e^{-\frac{\tilde x^2+\tilde y^2}{2}},
\end{aligned}
\tag{S14}
\label{WF1}
\end{equation}
with $\gamma_b(t) = -\frac{i}{2}\tanh(\zeta t)$.
In the first line of Eq.~(\ref{WF1}), we simplify the squeezing operator using the Baker-Campbell-Hausdorff (BCH) formula, i.e., \cite{Fisher1987}
\begin{equation}
\begin{aligned}
e^{-i \frac{\zeta t}{2} [(\tilde b^\dagger)^2  + \tilde b^2]} |0,0 \rangle &=  e^{-\frac{i}{2}\tanh(\zeta t) (\tilde b^\dagger)^2}  e^{-\ln[\cosh(\zeta t)]\left(\tilde b^{\dagger} \tilde b+\frac{1}{2}\right)}  e^{-\frac{i}{2}\tanh(\zeta t) b^2} |0,0 \rangle \\
&=  e^{\gamma_b(t) (\tilde b^\dagger)^2}  e^{-\frac{1}{2}\ln[\cosh(\zeta t)]} |0,0 \rangle \\
&=  \frac{1}{\sqrt{\cosh(\zeta t)}} e^{\gamma_b(t) (\tilde b^\dagger)^2} |0,0 \rangle.
\end{aligned}
\tag{S15}
\end{equation}
In the second line of Eq.~(\ref{WF1}), we used the Eq.~(\ref{zDz}) along with the commutator outlined in Eq.~(\ref{zDzCommut}). Based on Eq.~(\ref{WF1}), the density distribution can be obtained as
\begin{equation}
\begin{aligned}
\rho(\mathbf{\tilde r},t) = |\langle \tilde x,\tilde y|0,S(t)\rangle|^2 &= \frac{1}{\pi \cosh (\zeta t)} e^{-[\tilde x^2+\tilde y^2+2\tanh(t\zeta) \tilde x\tilde y]} \\
&= \frac{1}{\pi \cosh (\zeta t)} e^ {-[1-\tanh (\zeta t)] \frac{( \tilde x+ \tilde y)^2}{2}-[1+\tanh (\zeta t)] \frac{( \tilde x- \tilde y)^2}{2}},
\end{aligned}
\tag{S16}
\label{rho2}
\end{equation}
namely Eq.~(9) of the main text.

\item \textbf{Cyclotron mode squeezing.} The stroboscopic wave function of the cyclotron $\tilde a$-mode squeezed state can be derived as
\begin{equation}
\begin{aligned}
\langle \tilde x,\tilde y|S(t),0\rangle &= \langle \tilde x,\tilde y| e^{i \frac{\zeta' t}{2} [(e^{i \frac{\varphi}{2}}\tilde{a}^{\dagger})^2 + (e^{-i \frac{\varphi}{2}}\tilde{a})^2]} |0,0 \rangle = \langle \tilde x,\tilde y| \frac{1}{\sqrt{\cosh(\zeta' t)}} e^{\frac{i}{2}e^{i \varphi} \tanh(\zeta' t) (\tilde a^\dagger)^2 } |0,0 \rangle \\
&=  \langle \tilde x,\tilde y| \frac{1}{\sqrt{\cosh(\zeta' t)}} e^{\gamma_a(t) (\bar z- \tilde b)^2 } |0,0 \rangle  = \langle \tilde x,\tilde y| \frac{1}{\sqrt{\cosh(\zeta' t)}} e^{\gamma_a(t) \bar{z}^2} |0,0 \rangle \\
&= \frac{1}{\sqrt{\cosh(\zeta' t)}} e^{\gamma_a(t) \bar{z}^2 } \langle \tilde x,\tilde y|0,0\rangle \\
&= \frac{1}{\sqrt{\pi\cosh(\zeta' t)}} e^{\gamma_a(t) (\tilde x-i\tilde y)^2 } e^{-\frac{\tilde x^2+\tilde y^2}{2}},
\end{aligned}
\tag{S17}
\label{WF2}
\end{equation}
with $\gamma_a(t) = \frac{i}{2}e^{i \varphi} \tanh(\zeta' t)$.
Accordingly, the density distribution can be derived as
\begin{equation}
\rho(\mathbf{\tilde r},t) = |\langle \tilde x,\tilde y|S(t),0\rangle|^2 = \frac{1}{\pi \cosh (\zeta' t)}
e^ {-(\tilde x^2+\tilde y^2)-\tanh (\zeta' t)\left[\sin (\varphi)\left(\tilde x^2-\tilde y^2\right)- 2 \cos (\varphi) \tilde x \tilde y\right]}.
\tag{S18}
\label{rho2}
\end{equation}
For $t_0 = 0$ and $\varphi = 0$, $\rho(\mathbf{\tilde r},t)$ can be reduced to Eq.~(15) of the main text; whereas for $t_0 = T/4$ and $\varphi = -\nu t_0 = -\pi$, $\rho(\mathbf{\tilde r},t)$ behaves as
\begin{equation}
\rho(\mathbf{\tilde r},t) = \frac{1}{\pi \cosh (\zeta' t)}e^ {-[1+\tanh (\zeta' t)] \frac{( \tilde x+ \tilde y)^2}{2}-[1-\tanh (\zeta' t)] \frac{( \tilde x- \tilde y)^2}{2}}.
\tag{S19}
\label{rho2}
\end{equation}

\item \textbf{Two-mode squeezing.} We now consider the most general case of the two-mode geometric squeezing with $\zeta' \neq \zeta$. The wave function in the coordinate space can be derived by
 \begin{equation}
\begin{aligned}
\langle \tilde x,\tilde y|S'(t),S(t)\rangle &= \langle \tilde x,\tilde y| e^{i \frac{\zeta' t}{2} [(e^{i \frac{\varphi}{2}}\tilde{a}^{\dagger})^2 + (e^{-i \frac{\varphi}{2}}\tilde{a})^2]} e^{-i \frac{\zeta t}{2} [(\tilde b^\dagger)^2  + \tilde b^2]} |0,0 \rangle \\
&=\frac{1}{\sqrt{\cosh(\zeta' t)\cosh(\zeta t)}} \langle \tilde x,\tilde y|  e^{\gamma_a(t) (\tilde a^\dagger)^2 } e^{\gamma_b(t) (\tilde b^\dagger)^2 } |0,0 \rangle \\
&=\frac{1}{\sqrt{\cosh(\zeta' t)\cosh(\zeta t)}} \langle \tilde x,\tilde y|  e^{\gamma_a(t) (\tilde a^\dagger)^2 } e^{\gamma_b(t) z^2 } |0,0 \rangle \\
&=\frac{1}{\sqrt{\cosh(\zeta' t)\cosh(\zeta t)}} \langle \tilde x,\tilde y|  e^{\gamma_a(t)\left( \frac{\bar{z}}{2} - \partial \right)^2 } e^{4\gamma_b(t) (a+ \bar \partial)^2 } |0,0 \rangle \\
&=\frac{1}{\sqrt{\cosh(\zeta' t)\cosh(\zeta t)}} \iint \diff \tilde p_x \diff \tilde p_y \langle \tilde x,\tilde y|  e^{\gamma_a(t)\left( \frac{\bar{z}}{2} - \partial \right)^2 }|\tilde p_x,\tilde p_y\rangle\langle \tilde p_x,\tilde p_y| e^{4\gamma_b(t) \bar \partial ^2 } |0,0 \rangle \\
&=\frac{1}{\sqrt{\pi\cosh(\zeta' t)\cosh(\zeta t)}} \iint \diff \tilde p_x \diff \tilde p_y  e^{\gamma_a(t)\left( \frac{\bar{z}}{2} - \partial \right)^2 }e^{4\gamma_b(t) \bar \partial ^2 } \langle \tilde x,\tilde y|\tilde p_x,\tilde p_y\rangle\langle \tilde p_x,\tilde p_y|0,0 \rangle \\
&=\frac{1}{\sqrt{\pi\cosh(\zeta' t)\cosh(\zeta t)}} \iint \frac{\diff \tilde p_x \diff \tilde p_y}{2\pi}  e^{\gamma_a(t)\left( \frac{\bar{z}}{2} - \partial \right)^2 }e^{4\gamma_b(t) \bar \partial ^2 } e^{i (\tilde p_x \tilde x+\tilde p_y \tilde y)} e^{-\frac{\tilde p_x^2+\tilde p_y^2}{2}} \\
&=\frac{1}{\sqrt{\pi(1-4 \gamma_a \gamma_b) \cosh (\zeta' t) \cosh (\zeta t)}}e^{\frac{[(2 \gamma_a-1) \tilde x-i(2 \gamma_a +1)\tilde y][(2 \gamma_b-1) \tilde x+i(2 \gamma_b+1)\tilde y]}{8 \gamma_a \gamma_b-2}}
\end{aligned}
\tag{S20}
\label{WF3}
\end{equation}
In the 4th line, we have adopted the substitutions displayed in Eqs.~(\ref{abz}) and (\ref{zDz}). We introduced the complete basis of momenta, i.e., $|\tilde p_x,\tilde p_y\rangle$, in the 5th line, and used the commutators shown in Eq.~(\ref{commuz}) to obtain the 6th line. 
Although the general expression given in Eq.~(\ref{WF3}) is quite complicated, it can easily reproduce the results of single-mode squeezing as demonstrated in Eq.~(\ref{WF1}) and Eq.~(\ref{WF2}) by setting $\zeta' = 0$ and $\zeta = 0$, respectively. 

In the case of $\zeta' = \zeta$, Eq.~(\ref{WF3}) can be simplified into
\begin{equation}
\langle \tilde x,\tilde y|S(t),S(t)\rangle = \frac{e^{\frac{[\tanh (\zeta t)(\tilde x+i \tilde y) -i \tilde x-\tilde y][\tanh (\zeta t)e^{i \varphi}(\tilde x-i \tilde y) +i \tilde x-\tilde y]}{-2+2 e^{i \varphi} \tanh ^2(\zeta t)}}}{[\pi(\cosh ^2(\zeta t)-e^{i \varphi} \sinh ^2(\zeta t))]^{1/2}},
\label{WF4}
\tag{S21}
\end{equation}
For the stroboscopic dynamics at moments $t = nT$ with $t_0 = \varphi = 0$, the wave function can be further reduced to
\begin{equation}
\begin{aligned}
\langle \tilde x,\tilde y|S(t),S(t)\rangle &= e^{-\frac{1}{2}\left(\tilde x^2+\tilde y^2\right) \cosh (2 \zeta t)+\tilde x \tilde y \sinh (2 \zeta t)} \\
&= \frac{1}{\sqrt \pi}e^{-\frac{1}{4} e^{-2 \zeta t}(\tilde x+\tilde y)^2-\frac{1}{4} e^{2 \zeta t}(\tilde x-\tilde y)^2},
\end{aligned}
\tag{S22}
\end{equation}
with the density distribution
\begin{equation}
\begin{aligned}
\rho(\mathbf{\tilde r},t) = \frac{1}{\pi}e^{-\frac{1}{2} e^{-2 \zeta t}(\tilde x+\tilde y)^2-\frac{1}{2} e^{2 \zeta t}(\tilde x-\tilde y)^2},
\end{aligned}
\tag{S23}
\end{equation}
which reproduces Eq.~(17) of the main text. For the stroboscopic dynamics in quarter periods with $t_0 = T/4$ and $\varphi = -\pi$, we have 
\begin{equation}
\langle \tilde x,\tilde y|S(t),S(t)\rangle = \frac{e^ {-\frac{1}{2}\left(\tilde x^2+\tilde y^2\right) \operatorname{sech}(2 \zeta t)-\frac{i}{2} \left(\tilde x^2-\tilde y^2\right) \tanh (2 \zeta t)}}{\sqrt{\pi \cosh (2 \zeta t)}},
\tag{S24}
\end{equation}
and the corresponding density distribution
\begin{equation}
\rho(\mathbf{\tilde r},t) = \frac{e^{-\frac{\tilde x^2+\tilde y^2}{\cosh(2 \zeta t)}}}{\pi \cosh (2 \zeta t)},
\tag{S25}
\end{equation}
with the latter reproducing Eq.~(18) of the main text.

\end{itemize}

\begin{figure}[t]
	 \includegraphics[width=0.99\textwidth]{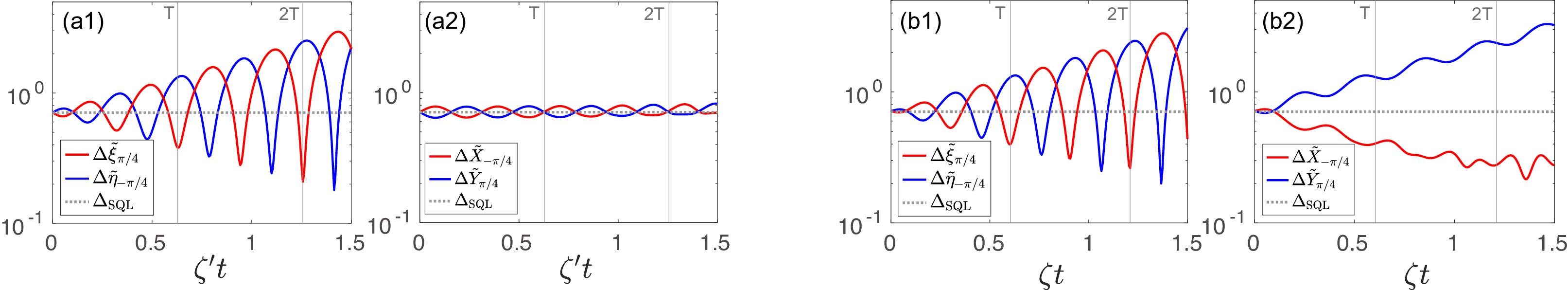}
	\caption{(a) Squeezing dynamics for the single cyclotron mode squeezing with $\varepsilon=0$, $\varepsilon'=0.4$ and $\Omega = \omega$, where panels (a1) and (a2) show the quadrature fluctuations in the $\tilde \xi$-$\tilde \eta$ and $\tilde X$-$\tilde Y$ phase spaces, respectively. (b) Squeezing dynamics for the two-mode squeezing with $\varepsilon = \varepsilon'=0.4$ and $\Omega \approx 1.02\omega$.}
	\label{FigS1}
\end{figure}

\section{Floquet Geometric Squeezing for relatively large $\varepsilon$}
\label{SecIII}
In Fig.~2 of the main text, we demonstrated the Floquet protocol for non-interacting BECs under a small trapping anisotropy $\varepsilon(t)$. Here, we show our protocol can also accommodate non-interacting BECs with larger values of $\varepsilon(t)$. According to Eqs.~(8) and (14) of the main text, a larger $\varepsilon(t)$ (namely a larger $\zeta$ or $\zeta'$) would lead to higher efficiency for both the squeezing and the anti-squeezing.

For single cyclotron mode squeezing with $\varepsilon = 0$ and $\varepsilon' \neq 0$, we have $s = 1$, $\zeta ' =  \varepsilon' \omega/2$, and $\nu = 2 \tilde{\omega}_+ = 4\omega$. In Fig.~\ref{FigS1}(a), we display the squeezing dynamics of the cyclotron mode with $\varepsilon' = 0.4$. Panels (a1) and (b2) show the quadrature fluctuations in the $\tilde \xi$-$\tilde \eta$ and $\tilde X$-$\tilde Y$ phase spaces, respectively. The results exhibit quite similar behaviors to those observed in the case of $\varepsilon' = 0.2$ [see Figs.~2(a1) and (a2) of the main text]. Compared to the latter, squeezing at $\varepsilon' = 0.4$ is more efficient, as evidenced by the greater amplitude of squeezing and anti-squeezing achieved over one stroboscopic period $T$. Note that $T = 2\pi/\nu = \pi/2\omega$, which is independent on $\varepsilon'$.

For the two-mode squeezing with $\varepsilon = \varepsilon' \neq 0$, we have $s = \sqrt{1+(\varepsilon \omega/2\Omega)^2}$ and $\tilde{\omega}_- = s\omega - \Omega$. In this situation, the influence of the $\tilde{\omega}_- \tilde{b}^{\dagger} \tilde{b}$ term in $\tilde{h}_0^\text{eff}$ [Eq.~(13) of the main text] can no longer be neglected. We find that, when 
\begin{equation}
\Omega = \sqrt{\frac{1+\sqrt{1+\varepsilon^2}}{2}}\omega,
\tag{S26}
\label{OmegaR}
\end{equation}
 $\omega_-$ vanishes and $\zeta = \zeta' = \frac{\varepsilon \omega}{\sqrt{1+\sqrt{1+\varepsilon^2}}}$. In Fig.~\ref{FigS1}(b), we display the dynamics of the two-mode squeezing for $\varepsilon = \varepsilon' = 0.4$, where $\Omega \approx 1.02\omega$ according to Eq.~(\ref{OmegaR}). The results are qualitatively consistent with those observed for $\varepsilon = \varepsilon' = 0.2$ [see Figs.~2(b1) and (b2) of the main text].

\section{High-Order Correction to Floquet Effective Hamiltonian}
\label{SecIV}
The Floquet effective Hamiltonian $\tilde{h}_0^\text{eff}$ [Eq.~(13) in the main text] serves as the leading-order approximation of the periodically driven Hamiltonian $\tilde{h}_0$ [Eq.~(12)]. Here, we calculate the second-order correction based on the high-frequency expansion. 

The second-order correction $H_F^{(2)}$ is generally given by \cite{Eckardt2015}
\begin{equation}
H_F^{(2)} = \sum_{m>0} \frac{[H_m, H_{-m}]}{m\nu},
\tag{S27}
\label{HF2}
\end{equation}
where $H_m$ is the Hamiltonian at the $m$-th Fourier component. Based on Eq.~(12) of the main text, we have
The Fourier components $H_m$ are:

\begin{equation}
\begin{aligned}
H_0 &= \tilde \omega_- \tilde{b}^{\dagger} \tilde{b} - \frac{\zeta'}{2} \tilde{a}^2 - \frac{\zeta'}{2} \tilde{a}^{\dagger 2} - \frac{\zeta'}{2} \tilde{b}^2 - \frac{\zeta'}{2} \tilde{b}^{\dagger 2}, \\
H_{1} &= -\zeta' \tilde{a} \tilde{b}^\dagger,\ \ H_{-1} =  -\zeta' \tilde{a}^\dagger \tilde{b}, \\
H_{2} &= -\frac{\zeta}{2} \tilde{a}^2 - \frac{\zeta}{2} \tilde{b}^2,\ \ H_{-2} = -\frac{\zeta}{2} \tilde{a}^{\dagger 2} - \frac{\zeta}{2} \tilde{b}^{\dagger 2}, \\
H_{3} &= -\zeta' \tilde{a}^\dagger \tilde{b},\ \ H_{-3} = -\zeta' \tilde{a} \tilde{b}^\dagger, \\
H_{4} &= -\frac{\zeta'}{2} \tilde{a}^2,\ \ H_{-4} = -\frac{\zeta'}{2} \tilde{a}^{\dagger 2}.
\end{aligned}
\tag{S28}
\label{HFF}
\end{equation}
Then, a straightforward calculation yields:
\begin{equation}
\begin{aligned}
H_F^{(2)} = \frac{-5(\zeta')^2 + 6\zeta^2}{12\nu} \tilde{a}^\dagger \tilde{a} + \frac{4(\zeta')^2 + 3\zeta^2}{6\nu} \tilde{b}^\dagger \tilde{b} + \text{const}.
\end{aligned}
\tag{S29}
\label{HF2_2}
\end{equation}
For the two-mode squeezing case where $\zeta' = \zeta$ and $\nu = 2\tilde{\omega}_+$, Eq.~(\ref{HF2_2}) simplifies to:
\begin{equation}
H_F^{(2)} = \frac{\zeta^2}{24\tilde{\omega}_+} \tilde{a}^\dagger \tilde{a} + \frac{7\zeta^2}{12\tilde{\omega}_+} \tilde{b}^\dagger \tilde{b} + \text{const}.
\tag{S30}
\end{equation}